\documentclass[aip,jcp,reprint,twocolumn,amsmath,amssymb,amsfonts,floatfix,letterpaper,longbibliography]{revtex4-2}
\usepackage{mathptmx} 
\usepackage{courier} 
\normalfont
\usepackage[T1]{fontenc}
\usepackage[flushleft]{threeparttable}
\usepackage{multirow}
\usepackage{dcolumn}
\usepackage{bm}
\usepackage{graphicx}
\usepackage[version=3]{mhchem}
\usepackage{braket}
\usepackage{color}
\usepackage{booktabs}
\usepackage{enumerate}
\usepackage{ulem}
\usepackage[font=normalsize]{caption}

\renewcommand{\phi}{\varphi}
\renewcommand{\epsilon}{\varepsilon}

\begin{document}

\title{Orbital choice in constructing model Hamiltonians}
\date{\today}
\author{David Wilian Oliveira de Sousa}
\affiliation{Department of Chemistry, Pennsylvania State University, University Park, Pennsylvania 16802, USA}
\author{Elvira R. Sayfutyarova}
\affiliation{Department of Chemistry, Pennsylvania State University, University Park, Pennsylvania 16802, USA}

\begin{abstract}
Model Hamiltonians (e.g., the Hubbard or Heisenberg models) provide a simple yet physically meaningful description of complex electronic phenomena through a small number of parameters, such as electron hopping integrals or magnetic exchange couplings. Their construction typically exploits the local nature of electron correlation effects and therefore relies on localized orbitals. Although many localization schemes are available, the resulting localized orbitals depend not only on the chosen localization functional but also on the orbital space to which the localization procedure is applied. These choices are often treated as technical details and are rarely discussed explicitly. Here, we demonstrate that different, chemically reasonable choices of localized orbitals can lead to substantially different model Hamiltonian parameters, even when they produce nearly identical electronic energies. Using Density Matrix Downfolding, we systematically investigate how the localization functional and, more importantly, the choice of model orbital space affect effective Hamiltonians derived from ab initio calculations for representative $\pi$-conjugated systems and transition metal complexes.
\end{abstract}

\maketitle

\section{Introduction}
Model Hamiltonians proved to be very useful in physics and chemistry. The main advantage of model Hamiltonians over all-electron exact Hamiltonians is their simplicity and transparency. They are typically expressed through a small set of parameters that can provide a clear, human-understandable interpretation of various physical phenomena and establish a connection between experimental measurements and the underlying physical interactions. Model Hamiltonians are often capable of describing complex physics through simple mechanisms or properties that can be generalized for a whole class of molecules, while full quantum chemical calculations typically characterize only one molecule at a time. Finally, model Hamiltonians are much simpler than the \textit{ab initio} Hamiltonians used in quantum mechanics, allowing one to find their eigenstates and eigenvalues with lower computational cost. This opens up ways for using them in dynamic simulations, especially in cases where \textit{ab initio} molecular dynamics can be too expensive.

Below, we briefly discuss several model Hamiltonians that have been successfully used to describe various physical phenomena. The H{\"u}ckel model \cite{Huckel1,Huckel2, Huckel3,hoffmann1963extended} not only served as one of the first molecular orbital theories, but has been widely used to describe qualitatively physical phenomena arising from the interactions of $\pi$-electrons in molecules with many conjugated $\pi$-bonds. The Hubbard \cite{hubbard1963electron,hubbard1964electron1,hubbard1964electron2} model captures the rich physics of electron-electron interactions, ranging from insulating and magnetic to novel superconducting effects in condensed matter. It gained popularity after providing rationalization for Mott insulators (i.e., antiferromagnetic insulators, like transition metal oxides that are expected to exhibit metallic properties according to conventional band theories).\cite{mott1937discussion,mott1949basis,mott1968metal}
 The Heisenberg-Dirac-Van Vleck Hamiltonian \cite{Heisenberg,van1932theory} is a convenient model to describe antiferromagnetically and ferromagnetically coupled spin centers in terms of exchange coupling parameters. 

The parameters of such model Hamiltonians can be determined either from empirical data or from \textit{ab initio} calculations.  
 Since model Hamiltonians are typically developed to describe the low-energy space of the strongly correlated systems, i.e., the systems with (near-)degenerate ground and excited states, the electronic states are usually obtained from multiconfigurational and/or multireference methods, in which the electron correlation is partitioned into static (short-range part) and dynamic (long-range part) contributions.
Static correlation is related to the strong mixing of degenerate or near-degenerate configurations, and the set of these configurations constitutes the model space.
The orbitals used to generate all these configurations constitute the so-called active space.
Dynamic correlation ( i.e., correlation of electronic motion at short range to allow electrons to avoid each other), 
requires accounting for interactions between the active orbitals and all orbitals outside the active space. 
As the computational cost is growing very fast with the size of the system and computational basis set, dynamic correlation is often neglected. In such cases, the quality of the model space strongly influences the accuracy of the resulting model Hamiltonian, making the choice of active orbitals particularly important.


To extract the parameters of model Hamiltonians from electronic structure calculations, one must first choose the orbital basis in which the model Hamiltonian is represented. 
Since many model Hamiltonians describe local interactions between electrons on the same atom (or lattice site), or between neighboring atoms, it is rational to employ spatially localized orbitals (one-electron functions).
Depending on the type of model Hamiltonian, this basis may consist of atomic orbitals or localized molecular orbitals. For example, the H{\"u}ckel, Hubbard, and Pariser-Parr-Pople Hamiltonians are formulated in a basis of atomic-like valence orbitals (e.g., one $p$ orbital per atom in $\pi$-conjugated systems), whereas effective Hamiltonians for magnetic systems are commonly expressed in terms of localized molecular orbitals. Since computational basis sets generally do not contain basis functions corresponding directly to atomic orbitals ($s$-, $p$-, $d$-, etc.), the identification of atomic orbitals typically requires an auxiliary basis, such as the minimal basis set of tabulated free-atom atomic orbitals (MINAO). The intrinsic atomic orbitals (IAOs) \cite{knizia:iao} provide a convenient alternative, as they represent a minimal basis of core and valence atomic orbitals polarized by the molecular environment while exactly spanning the occupied SCF space.

An alternative approach is using compact molecular orbitals obtained with different orbital localization methods introduced in quantum chemistry. \cite{BenAmor2021} Due to the invariance of the SCF (Hartree-Fock or Kohn-Sham) wave function with respect to unitary transformations among the occupied molecular orbitals, 
many localization procedures can be used to construct localized orbitals that differ significantly in shape, structure, and locality while representing exactly the same $N$-electron wave function.

While the validity and accuracy of model Hamiltonians for describing specific systems, and physical phenomena have been extensively studied, little attention has been paid to the dependence of the model Hamiltonian parameters on the choice of orbital basis. 
Since the matrix elements of model Hamiltonians are evaluated in the chosen orbital basis, different choices of atomic or localized molecular orbitals generally lead to different Hamiltonian parameters, even though they describe the same underlying electronic wave function.
In this paper, we investigate how variations in the orbital basis impact the optimized parameters of model Hamiltonians extracted from full $N$-electron \textit{ab initio} calculations.

The model Hamiltonians  are derived using recently introduced density matrix downfolding (DMD)\cite{DMD1,DMD2} method, which allows model Hamiltonians of arbitrary form, while matching the low-energy spectrum of the \textit{ab initio} Hamiltonian to that of the model Hamiltonians.

\section{Methods}
\label{sec:methods}
In section \ref{sec:p1} we first briefly describe how to extract the effective Hamiltonians from \textit{ab initio} calculations using DMD.  For further details, we refer the reader to Ref.\cite{DMD1, DMD2}. In Section \ref{sec:p2} we describe the models tested in this work. In section \ref{sec:p3} we describe common orbital localization methods used in quantum mechanical calculations. 
In Section \ref{sec:p4} we summarize how one can utilize the orbital localization methods for systems with many conjugated $\pi$-bonds and transition metal complexes, 
 two most types of strongly correlated systems that are often studied using model Hamiltonians.
\subsection{Extracting effective Hamiltonians using DMD}
\label{sec:p1}

Consider a quantum system with the \textit{ab initio} nonrelativistic electronic Hamiltonian $\hat H$ in a $N$-electron Hilbert space $\mathcal{H}$:
\begin{align}
\hat H=\sum_{ij}t_{ij} \hat E_{ij} +\frac{1}{2}\sum_{ijkl} V_{ijkl} \hat e_{ij,kl}.  \label{eq:Ham}
\end{align}
Here $\hat E_{ij}=\sum_{\sigma}\hat c_{i\sigma}^\dagger \hat c_{j\sigma}$ and $\hat e_{ij,kl}=\hat E_{ij}\hat E_{kl}-\delta_{jk}\hat E_{il}=\sum_{\sigma, \tau}\hat c_{i\sigma}^\dagger \hat c_{k\tau}^\dagger \hat c_{l\tau}\hat c_{j\sigma}$ represent one- and two-electron excitation operators, defined in terms of the creation and annihilation operators $\hat c_{i\sigma}^\dagger$, $\hat c_{k\tau}^\dagger$  $ \hat c_{j\sigma}$ and $ \hat c_{l\tau}$ associated with spatial orbitals $i,j,k,l$ and spin $\sigma$ and $\tau$; $t_{ij}$ and $V_{ijkl} $ denote the one- and two-electron integrals in this spatial orbital basis.

For a wave function $\ket {\psi} \in \mathcal{H} $, the energy functional is given by:
 \begin{align}
  E[\psi]=\frac{\bra \psi \hat H \ket \psi }{\braket {\psi|\psi} } \label{eq:E_Ham}
\end{align}
and $\bra \psi \hat E_{ij} \ket \psi $ and $\bra \psi \hat e_{ij,kl} \ket \psi $ provide elements of the spin-free one- and two-electron reduced density matrices (RDMs).
The DMD method uses RDMs (instead of eigenstates utilized by other downfolding approaches) obtained for \textit{ab initio} wave functions in the low-energy space to extract a model Hamiltonian with a chosen set of parameters. 

Let us define the low-energy space $\mathcal{L} (\hat H,N)$ as a subspace of $\mathcal{H} $ spanned by $N$ vectors, which have the $N$ lowest eigenvalues of $\hat H$, $\{\ket \psi \}_n$, so that we have:
\begin{align}
\hat{H} \ket {\psi_n}=E_n \ket {\psi_n}.
\end{align}
The effective Hamiltonian, $\hat H_{\mathrm{eff}}$, is an operator on the Hilbert space  $\mathcal{L} $, that aims to approximate the \textit{ab initio}
Hamiltonian and in the sense of reprodicing its spectrum within the low-energy space $\mathcal{L}$. Its energy functional is given by:
 \begin{align}
  E_{\mathrm{eff}}[\psi]=\frac{\bra \psi \hat H_{\mathrm{eff}} \ket \psi }{\braket {\psi|\psi} } \label{eq:E_Ham_eff}.
\end{align}
The derivation of the optimal effective Hamiltonian with the DMD method\cite{DMD2} requires matching the effective energy functionals from Eq.~(\ref{eq:E_Ham_eff}) with those of the full \textit{ab initio} Hamiltonian from Eq.~(\ref{eq:E_Ham}) for all wavefunctions  $\ket {\psi_n} \in \mathcal{L} (\hat H,N)$ , i.e.
 $ E_{\mathrm{eff}}[\psi]= E[\psi]$ and $\hat{H}_{\mathrm{eff}}\ket {\psi_n}=\hat{H} \ket {\psi_n}$.
Thus, the general protocol of the DMD method includes obtaining \textit{ab initio} wavefunctions $\ket {\psi_n}$ in the low-energy space (sampling) and computing the effective energy functionals using a chosen parametrized form of $\hat{H}_{eff}$, then optimizing the parameters. The effective functional can be represented in terms of descriptors $d_r[\psi]$, that are either RDMs or properties that can be expressed through RDMs (e.g., spin-spin correlation):
\begin{align}
 E_{\mathrm{eff}}[\psi]=\sum_{r} f_r(d_r[\psi]),  \label{eq:E_via_ds}
\end{align}
where $f_r$ are some parameterized functions.

For example, if we assume the effective Hamiltonian of the form
\begin{align}
\hat H_{\mathrm{eff}}=E_0+\sum_{ij}t_{ij} \hat E_{ij}  +\sum_{ijkl} V_{ijkl} \hat e_{ij,kl},  \label{eq:Heff}
\end{align}
with parameters $E_0$, $t_{ij}$ and $V_{ijkl}$, then the expectation value for a wavefunction $\ket {\psi_n}$
\begin{align}
 E_{\mathrm{eff}}[\psi_n]=E_0+\sum_{ij}t_{ij} \bra {\psi_n}  \hat E_{ij}  \ket {\psi_n}+\sum_{ijkl}  V_{ijkl}\bra {\psi_n} \hat e_{ij,kl} \ket {\psi_n},
\end{align}
readily follows from the form of Eq.~(\ref{eq:E_via_ds}): the summation terms represent one-electron and two-electron RDM elements and the energy functional can be written as:
\begin{align}
 E_{\mathrm{eff}}[\psi_n]=\sum_{r}p_r d_r[\psi_n].
\end{align}
Reformulated into matrix form, this becomes :
 \begin{align}
\mathbf{E}=\mathbf{D}\mathbf{p}, \label{eq:EDp}
\end{align}
where $\mathbf{E}=(E_1, E_2, . . . , E_M)^T$ is the $M$-dimensional vector containing the sampled energies obtained for the \textit{ab initio} Hamiltonian, 
$\mathbf{D}$ is the $M \times N_p$ matrix composed of density matrix elements (that play the role of descriptors here), and $\mathbf{p}= (E_0, t_{ij} ...V_{ijkl})^T$ is a $N_p$-dimensional vector of parameters. For each of the $M$ sample wavefunctions $ \ket {\psi_n}$, the ab initio energies have to match the effective energies written in terms of the RDM elements, $\gamma_{ij,n}=\bra {\psi_n} \hat E_{ij}   \ket {\psi_n}$ and $\Gamma_{ijkl,n}=\bra {\psi_n}\hat e_{ij,kl} \ket {\psi_n}$:
 \begin{align}
\left ( \begin{matrix}
 E_{1} \\ 
 E_{2} \\ 
\vdots \\
E_{N}
\end{matrix} \right )=
\left ( \begin{matrix}
1&\gamma_{ij,1}&...&\Gamma_{ijkl,1}&...\\
1&\gamma_{ij,2}&...&\Gamma_{ijkl,2}&...\\
\vdots&\vdots&\vdots&\vdots&\vdots\\
1&\gamma_{ij,N}&...&\Gamma_{ijkl,N}&...
\end{matrix} \right )
\left ( \begin{matrix}
 E_{0} \\ 
t_{ij} \\ 
... \\
V_{ijkl} \\
...
\end{matrix} \right )
\end{align}

The optimization of parameters can be performed then, for example, using the least squares method, by minimizing the norm of the error $\Delta E$ of the effective energies with respect to the \textit{ab initio} energies:
\begin{align}
\Delta E=||\mathbf{E}-\mathbf{D}\mathbf{p}||^2=\sum^M_n \left (E_n- E_{eff}[\psi_n] \right)^2
\end{align}
In this case, the optimization problem is reduced to a linear least squares fit,  which can be solved in closed form via singular value decomposition.
It is also possible to consider other cost functions or effective Hamiltonians, including such for which the optimization would be nonlinear. 

In principle, the DMD method is exact in the following sense: it can construct an effective Hamiltonian $\hat H_{eff}$ that matches the full Hamiltonian $\hat H$ in its low-energy subspace always exists and is unique. 
In practice, the quality of the DMD is affected by the introduced approximations, such as the specific parameterized form of the effective Hamiltonian, and the quality of the energy states used to sample the low-energy space. The latter significantly depends on the electronic structure method chosen for calculations.  


\subsection{Model Hamiltonians used in this work}
\label{sec:p2}

The first model used in this work is the well-known Hubbard model. It contains two parameters, related to one-electron and two-electron interactions:
\begin{align}
\hat H_{m1}=E_0-t\sum_{i,\sigma} \left( \hat c_{i\sigma}^\dagger \hat c_{i+1\sigma} +\hat c_{i+1\sigma}^\dagger \hat c_{i\sigma} \right) +U\sum_{i}  \hat n_{i\uparrow} \hat  n_{i\downarrow},  \label{eq:MH1}
\end{align}
where $t$ is called the ``hopping integral'' between the nearest-neighbor orbitals (called ``sites'') , 
$U$ is the so-called on-site Coulomb repulsion (i.e., the repulsion between two electrons occupying the same orbital), and $\hat  n_{i\sigma}=\hat c_{i \sigma}^\dagger \hat c_{i \sigma}$. 
The negative sign in $t$ is conventionally chosen to keep the parameter positive. A zero-order parameter ($E_0$) is included to account for the ``core'' energy (the part of electronic energy not associated with the effective particles in the active space). 
In comparison with Eq.~(\ref{eq:Heff}), it considers only the case $j = i \pm 1$ (nearest neighbor) for the one-electron interactions, and only the case $i = j = k = l$ (two electrons in the same orbital) for the two-electron interactions. 
The nearest-neighbor approximation is used because normally the Hubbard model deals with lattice systems, where the orbitals are situated in a sequential or periodic pattern. It can also be used to model the $\pi$-space of organic conjugated systems or aromatic rings.

The second model considered in this work is intended to deal with molecules containing metal atoms (complexes). Its general form is given by:
\begin{align}
\hat H_{m2}=E_0&+\sum_{i} \epsilon_i n_i+ \sum_{ij, \sigma}t_{ij} \left( \hat c_{i \sigma}^\dagger \hat c_{j \sigma} + \hat c_{j \sigma}^\dagger \hat c_{i \sigma}\right)  \notag \\ 
&+U \sum_{i}  \hat n_{i \uparrow} \hat n_{i \downarrow} +\sum_{ij} V_{ij} \hat  n_{i}\hat  n_{j}+\sum_{i,j}K_{i,j}\hat {\mathbf{S}}_i \hat {\cdot \mathbf{S}}_j ,  \label{eq:MH3}
\end{align}
This model, unlike the previous ones, includes orbitals of different types, or from different atoms, so the orbital energies $\epsilon_i $ cannot be disregarded. The $t_{ij}$ and $V_{ij}$ terms can be grouped by the  kind of orbital or orbital symmetry. $K_{ij}$ is an exchange coupling describing magnetic interactions. $\hat {\mathbf{S}}_i \cdot \hat { \mathbf{S}}_j$ is the spin-spin correlation, expressed via local spin operators $\hat {\mathbf{S}}_i$ and $\hat {\mathbf{S}}_j$ acting on orbitals $i$ and $j$,  typically on semi-occupied orbitals (for example, the 3d orbitals of a first-row transition metal atom). 

The spin-spin correlation, $\hat {\mathbf{S}}_i \cdot \hat { \mathbf{S}}_j $  can also be written in terms of the creation and annihilation operators $\hat c_{i\sigma}^\dagger$ and $ \hat c_{j\sigma}$ (and, in turn, in terms of RDM elements):

\begin{align}
\hat {\mathbf{S}}_i \cdot \hat { \mathbf{S}}_j &=\hat {S}^x_i \hat {S}^x_j +\hat {S}^y_i \hat {S}^y_j+\hat {S}^z_i \hat {S}^z_j=\notag \\
&=\frac{1}{4} \left (\hat c_{i \uparrow}^\dagger \hat c_{i \uparrow}-\hat c_{i \downarrow}^\dagger \hat c_{i \downarrow} \right) \left (\hat c_{j \uparrow}^\dagger \hat c_{j \uparrow}- \hat c_{j \downarrow}^\dagger \hat c_{j \downarrow} \right )+ \notag \\ 
&+\frac{1}{2} \left (\hat c_{i \uparrow}^\dagger \hat c_{i \downarrow}\hat c_{j \downarrow}^\dagger \hat c_{j \uparrow}+\hat c_{i \downarrow}^\dagger \hat c_{i \uparrow}\hat c_{j \uparrow}^\dagger \hat c_{j \downarrow}  \right) =\notag \\
&=\frac{1}{4} \left (\hat n_{i \uparrow} -\hat n_{i \downarrow}  \right) \left (\hat n_{j \uparrow} -\hat n_{j \downarrow}  \right) -\frac{1}{2} \left(\hat e_{ij,ji} +\sum_\sigma  \hat n_{i \sigma} \hat n_{j \sigma} \right) \notag \\
&=-\frac{1}{2} \left( \hat e_{ij,ji} +\frac{\hat e_{ii,jj}}{2} \right) ,  \label{eq:SiSj_1}
\end{align}

In this work, we considered the benzene and naphthalene molecules as the case studies for derivation of the Hubbard model parameters with DMD. As benzene was considered in the original work on the DMD method,\cite{DMD1} it served as a benchmark for our own version of DMD. The Hubbard model for the $\pi$-system for both benzene and naphthalene has the same three parameters ($E_0$, $t$, and $U$). We used the FeSe molecule as an example of a transition metal complex, that can be described with the model Hamiltonian from Eq.~(\ref{eq:MH3}).


\subsection{Localized Molecular Orbitals}
\label{sec:p3}

Localized orbitals are widely used in quantum chemistry to describe and interpret molecular structure and chemical bonding, reflecting the local nature of electron correlation effects.  Despite the long history of orbital localization, there are not many conceptually different localization methods, and even of those,  not all of them are suitable for deriving model Hamiltonians. Here, we recapitulate some widely used  localization approaches for molecular orbitals: Edmiston-Ruedenberg (ER),\cite{edmiston:LocalizedAtomicAndMolecularOrbitals} Foster-Boys (FB), \cite{FB1, FB2}, and Pipek-Mezey (PM)\cite{pipek:PMlocalization}  methods.

The FB method aims to construct the most compact orbitals by directly minimizing their spatial extent through the following functional:
\begin{align}
   L_{\mathrm{FB}}   =& \sum_{i=1}^{N_\mathrm{orb}} \braket{\phi_{i}\phi_{i}| \left|\hat{\mathbf{ r}}_1-\hat{\mathbf{ r}}_2 \right|^2|\phi_{i}\phi_{i}}
   \label{eq:LocFB}
\end{align}

In solid-state physics, the FB functional was extended by Marzari and Vanderbilt to determine maximally localized Wannier functions.\cite{marzari1997maximally,marzari2012maximally} While the localization criterion is analogous to that of the molecular FB method, its formulation accounts for periodic boundary conditions and is expressed in terms of Bloch functions. A different real-space formulation of FB localization for periodic orbitals was also introduced by Zicovich-Wilson.\cite{zicovich2001general}


The ER method is based on the maximization of Coulomb self-repulsion:
\begin{align}
   L_{\mathrm{ER}}   =& \sum_{i=1}^{N_\mathrm{orb}} \braket{\phi_{i}\phi_{i}|  \frac{1}{\left|\hat{\mathbf{ r}}_1-\hat{\mathbf{ r}}_2 \right|}|\phi_{i}\phi_{i}}
   \label{eq:LocER}
\end{align}

The Pipek-Mezey method maximizes the sum of the squares of the partial atomic charges:
\begin{align}
   L_{\mathrm{PM}}   =& \sum_{A=1}^{N_\mathrm{Atoms}}\sum_{i=1}^{N_\mathrm{orb}} \left | Q^A_i\right|^2,
   \label{eq:LocPM}
\end{align}
where $Q^A_i$ is the partial charge associated with orbital $i$ on atom $A$. 

The PM localization depends significantly on the atomic orbitals used for computing partial charges. 
The original PM localization procedure relied on Mulliken charges, that are physically ill-defined and are very sensitive to the choice of the computational basis. L{\"o}wdin charges offer some improvement (e.g., unlike Mulliken populations, L{\"o}wdin populations cannot be negative),\cite{h2013pipek} but they still retain a significant dependence on the computational basis..
Using intrinsic atomic orbitals (IAOs) as the basis for computing partial atomic charges leads to intrinsic bond orbitals (IBOs)\cite{knizia2015electron}
that can be viewed as a special case of PM localization but without the dependence on the employed basis set. Recently, IBOs were used 
for the construction of plane-wave based Wannier functions \cite{schafer2021surface}.

Here we also use the recently introduced Nuclear Potential Localization (NPL) method,\cite{knizia2026} that maximizes the potential energy of the interaction between the orbital electron densities and the nuclei:
\begin{align}
   L_{\mathrm{NPL}}
=& \sum_{A=1}^{N_\mathrm{Nuc}}\sum_{i=1}^{N_\mathrm{orb}} h \left (\braket{\phi_{i}|  \frac{Z_A}{\left|\hat{\mathbf{ r}}-\hat{\mathbf{ R}}_A \right|}|\phi_{i}} \right),
   \label{eq:LocFnNpl}
\end{align}
where $h$ is a convex function (for further details, please, see Ref. \cite{knizia2026}). Here, we use $h(x)=x^2$.

These localization methods are most commonly applied to occupied orbitals. However, the same localization functionals can also be used within the virtual orbital space or within selected occupied or virtual subspaces. 
This means that there are many ways of constructing localized molecular orbitals even based on the same functional (e.g., the PM functional), and these choices typically depend on the system of interest and the computed properties.

\subsection{Orbital choices for $\pi$-conjugated systems and transition metal complexes}
\label{sec:p4}

The electronic structure of planar polycyclic aromatic hydrocarbons and other systems with many conjugated $\pi$ bonds typically features a set of $\pi$-orbitals,
formed by the $p$ atomic orbitals oriented perpendicular to the plane containing atomic nuclei.
Model Hamiltonians developed for $\pi$-conjugated systems (e.g., H{\"u}ckel, Hubbard, and Pariser-Parr-Pople) are formulated in a basis of one $p$-like orbital per atomic site. The resulting model space therefore corresponds to the complete $\pi$-subspace, containing both occupied and unoccupied $\pi$-orbitals of a system.

However, not every type of orbital localization method can separate molecular $\pi$-orbitals from $\sigma$-orbitals.
The ER and FB localization methods do not preserve  $\sigma$- and $\pi$-orbitals separation. Instead, the ER and FB localizations typically  result in the so-called ``banana bonds'', and therefore they are not directly applicable in their original form for such model Hamiltonians. 
However, it is possible to obtain localized $\pi$ orbitals using ER and FB localization functionals if one applies them exclusively on the $\pi$-orbitals subspace, the occupied and unoccupied canonical $\pi$-orbitals from the SCF calculation. In this case, they convert the canonical $\pi$-orbitals into localized $\pi$-orbitals.

The PM and IBO localization schemes preserve the separation between $\sigma$- and $\pi$-orbitals. 
However, the resulting localized $\pi$-orbitals \textit{depend on the orbital space to which the localization procedure is applied}. 

It should be noted that, although PM localization may formally be performed in the complete set of SCF canonical orbitals  (occupied+virtual), this approach is rarely used in practice because the resulting occupied orbitals no longer represent the original SCF wave function, i.e., occupied-virtual rotations change the underlying SCF wave function. Therefore, localization is typically performed within carefully chosen orbital subspaces rather than over the entire SCF orbital space. 

Starting from the canonical orbitals obtained from a Hartree-Fock or Kohn-Sham calculation, several practical choices of localization space are possible.

\begin{itemize}
   \item PM localization applied separately to the occupied and virtual orbital spaces. While this yields localized occupied and virtual orbitals, the resulting occupied $\pi$-orbitals generally remain delocalized over multiple atoms and are therefore not suitable as single atom-centered (or site) orbitals for constructing Hubbard or H{\"u}ckel-like Hamiltonians. In addition, localization of the complete virtual space typically does not yield physically meaningful localized orbitals unless the computation is performed in a minimal basis set \cite{subotnik2005fast} or unless additional transformations using an auxiliary minimal basis set are involved (e.g., see VVO \cite{schmidt2015valence}). In practice, localization of virtual orbitals yields chemically meaningful results only when it is restricted to the \textit{valence} virtual space.
 Localization of the full virtual space is also impractical on large systems, as the number of possible unitary orbital rotations grows rapidly with the size of the virtual space, often leading to numerous nearly equivalent solutions and  convergence issues.

   \item PM localization performed on the subset of manually selected occupied and virtual  canonical $\pi$-orbitals from the SCF calculations ($\pi$-subspace).   To produce highly localized $\pi$-orbitals resembling atomic $p$-orbitals, one should include the corresponding antibonding virtual orbitals in the localization space.\cite{thygesen2005partly} 
The inclusion of the virtual $\pi$-orbitals allows the bonding and antibonding combinations to rotate into orbitals localized on individual atomic sites. Without the corresponding antibonding orbitals, such a transformation is generally not possible.
This consideration is not limited to $\pi$-orbitals but applies generally to the localization of bonding orbitals on individual atomic centers. 

   \item PM localization performed on the subset of  $\pi$-orbitals constructed automatically using the $\pi$-orbital space (PiOS) method. \cite{sayfutyarova2019constructing} by projecting the canonical SCF orbitals onto the ``target $\pi$-space'' spanned by the atomic valence $p$ orbitals. Compared to manual selection of canonical $\pi$-orbitals, this approach avoids visual inspection and provides a reproducible and automated construction of the localization $\pi$-space.
The main advantage of this approach is that it is a well-defined procedure for generating $\pi$-orbitals localized on the desired atomic centers. None of other options do not explicitly enforce localization on a predefined set of atoms.

\end{itemize}

The same options are available for the IBO method or any other modified PM localization schemes utilizing atomic partial charges other than Mulliken charges.\cite{lehtola:GeneralizedPM}
The diversity of possible localization protocols is the reason why the details of the localization should always be provided.
Otherwise, reproducing or benchmarking results can be difficult when papers state that “PM localization” was used without specifying the details of its application.
Visualization alone does not  necessarily help, as localized orbitals obtained from different procedures may appear similar, especially when they are visualized using an isosurface threshold/cutoff corresponding to a small $\%$ of the orbital density.

For transition metal complexes (TMCs), the choice of orbital space depends on the model Hamiltonian and the properties of interest.

\begin{itemize}
   \item If only metal-centered $d$-like orbitals are required for modeling specific properties, canonical orbitals can sometimes be used directly, as the orbitals with predominantly metal $d$-orbital character are relatively compact in many TMCs.
Orbitals with dominant $d$-orbital character often correspond to singly occupied orbitals and can frequently be identified relatively easily.

\item ER and FB localization schemes can be used for many TMCs as they can produce compact orbitals on metal centers. However, obecause these localization schemes optimize real-space compactness, they mix different atomic orbital contributions on the same metal center when applied to the entire occupied orbital space. For example, for the first-row transition metals the localized molecular orbitals
contain $3s$, $3p$ and $3d$ atomic orbitals character, i.e., they are significantly hybridized (We provide an example in this work in section \ref{sec:tmcs}) . This can be a problem for modeling properties of TMCs related primarily to their valence $d$-orbitals.

In addition, these schemes may not be optimal for metal-carbonyl complexes, featuring the both $\sigma$ and $\pi$ metal-carbon bonds: the ligand-to-metal forward $\sigma$–donation and the metal-to-ligand $\pi$-backbonding.
In this case, one faces issues similar to those observed for polycyclic hydrocarbons or systems with many conjugated $\pi$ bonds, as the ER and FB localization methods do not separate  $\sigma$- and $\pi$-orbitals. 
\item Performance of the PM localization also depends on the type of metal-ligand bonding. It can also produce strongly mixed orbitals depending on the nature of the ligands.
\item All localization schemes can be applied to the subset of molecular orbitals, that is either manually selected or constructed automatically. One approach is to project the canonical SCF orbitals onto the ``target atomic valence-space'' spanned by the atomic valence $d$ orbitals on metal centers and, for example, ligand $p$ atomic orbitals of ligand atoms involved in metal-ligand bonding,  using the atomic valence active space (AVAS) method. \cite{sayfutyarova2017automated}.
As in the case of the PiOS method, the main advantage of this approach consists in avoiding  the visual inspection and manual selection of any orbitals. 
\item The recently introduced NPL method produces localized orbitals with properties similar to those obtained using the IBO method.
\end{itemize}


\section{Results and Discussion}
\label{sec:results}

For all systems, we first obtained the reference SCF wavefunction either from the Hartree-Fock or Kohn-Sham density functional theory calculations. Then so-computed canonical molecular orbitals were subjected to different localization schemes. 
To compute the ground and low-energy excited states, we used the complete active space configuration interaction (CASCI) method, which does not optimize or modify the orbitals. This allows us to see how the variations in the input localized orbitals impact the computed energy states and the model Hamiltonian parameters extracted for the low-energy space. Note that the same orbital set was used for every electronic state computed with the CASCI  method.
\begin{figure}[h!]
  \centering
  \includegraphics[width=1\columnwidth]{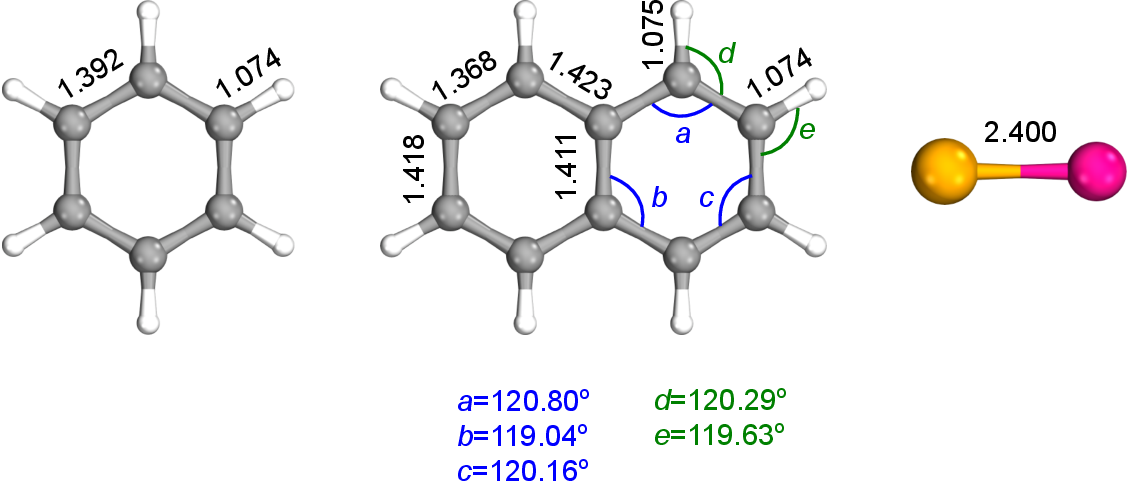}
  \caption{Geometries used in this work. All distances are given in {\r  A} (angstroms).}
  \label{fig:geom}
\end{figure}

All geometries used in this work are given in Figure \ref{fig:geom}. All SCF and CASCI calculations were performed using the PySCF package. ~\cite{sun2018pyscf,sun2020recent} 
The DMD fitting and the model Hamiltonian parameters extraction were performed using the Python-based code developed by the authors. The implemented DMD code is available on github.\cite{DMDus}

\subsection*{ Systems with many conjugated $\pi$-bonds }
\label{sec:pp}
Here we consider the Hubbard model for benzene and naphthalene as simple examples of molecular systems with many conjugated $\pi$-bonds. 
For both systems we obtained the SCF wavefunctions from the Hartree-Fock calculations using the def2-TZVP basis set.  \cite{Weigend:def2SVP_def2TZVPP} 
Benzene's $\pi$-system contains six $\pi$-electrons distributed among six $\pi$-orbitals, while in the case of naphthalene it contains ten $\pi$-electrons distributed among ten $\pi$-orbitals.
For benzene, we computed electronic states with S=0, 1, and 2, while for naphthalene we considered S=0 and 1. For each spin value S, we computed 20 electronic states.

When deriving the Hubbard model parameters for both benzene and naphthalene, we examined five different options: 
\begin{itemize}
\item A: PM applied to the subspace of canonical $\pi$ orbitals obtained from a SCF calculation
\item B: FB  applied to the subspace of canonical $\pi$ orbitals obtained from a SCF calculation
\item C: PM applied to the subspace of $\pi$ orbitals constructed with the PiOS method  
\item D: IAOs followed by selection of the $\pi$-orbital subspace.
 \end{itemize}

\begin{figure}[h!]
  \centering
  \includegraphics[width=1\columnwidth]{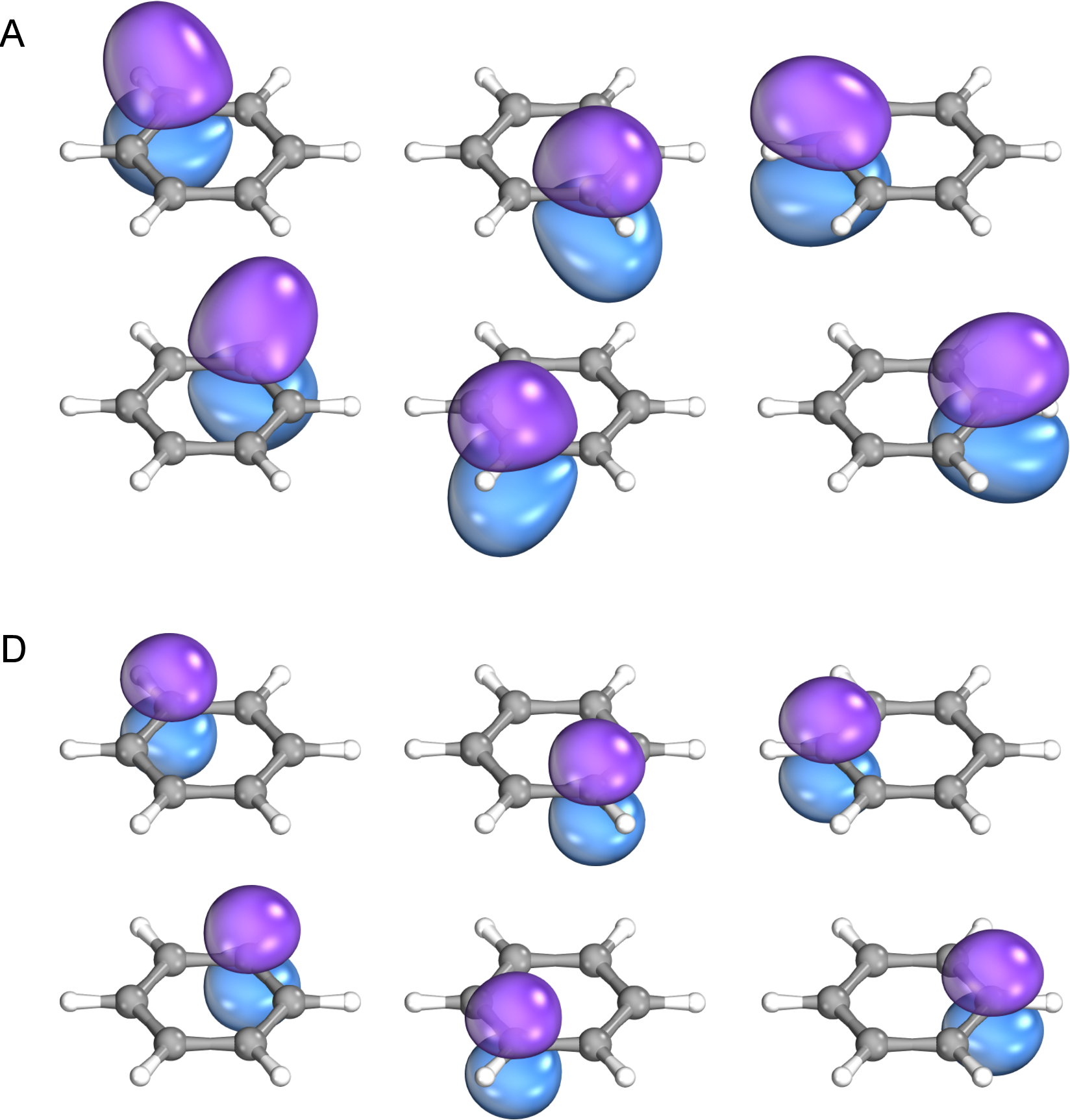}
  \caption{Localized $\pi$ orbitals obtained for benzene using three different options. Option A:  PM on a subset of canonical $\pi$-orbitals. Option D: IAOs. The orbitals are visualized with the isosurface threshold corresponding to an 80$\%$ density cutoff.}
  \label{fig:PM_C6H6}
\end{figure}

Visualization of orbitals obtained with these options gives very similar sets of localized orbitals centered on individual atoms and resembling atomic orbitals (see  Figure \ref{fig:PM_C6H6}).
However, different schemes for orbital localization produce different localized orbitals, even if they appear nearly indistinguishable during the visual inspection.  A detailed analysis of their composition in terms of atomic orbitals reveals that they exhibit different degrees of mixing with atomic orbitals located on other atomic centers. 

The differences in the resulting orbital spaces and orbital compositions manifest themselves in the computed energies and, more importantly, in the resulting model Hamiltonian parameters (Table \ref{tab:Hub_C6H6} and Table \ref{tab:Hub_C10H8}). 

\begin{table}[h!]
\caption{CASCI ground state energy and Hubbard model parameters for benzene, obtained with the DMD and CASCI electronic states}
\begin{center}
\label{tab:Hub_C6H6}
\begin{tabular}{ lccc}
\hline
Option& Ground state energy, a.u.&$U$, eV & $t$, eV \\
\hline
A&-230.825309& 3.67& 3.15  \\
B&-230.825309& 3.66 & 3.15 \\
C&-230.852771& 4.02 & 4.65 \\
D&-230.852771& 4.02 & 4.65\\
\hline
\end{tabular}
\end{center}
\end{table}

For benzene, options A and B produce practically identical ground-state energies because the localization was performed within the same (pre-selected) canonical $\pi$-orbital subspace.
The options C and D produce the same ground-state energies and the same model parameters. This happens because the PiOS method employs IAOs to target molecular orbitals with the desired atomic orbital character, and these two approaches generate orbitals spanning the same $\pi$-orbital subspace. 
The difference in CASCI ground-state energy between options A and C is about 0.7 eV,
but the differences in the computed on-site Coulomb repulsion and  hopping integrals are $\approx$ 0.4 eV and $\approx$ 1.5 eV, respectively (Table \ref{tab:Hub_C6H6}).
These results demonstrate that \textit{the choice of localization functional and the choice of orbital subspace represent two separate sources of variation in the extracted model parameters}.
 
\begin{figure}[h!]
  \centering
  \includegraphics[width=0.9\columnwidth]{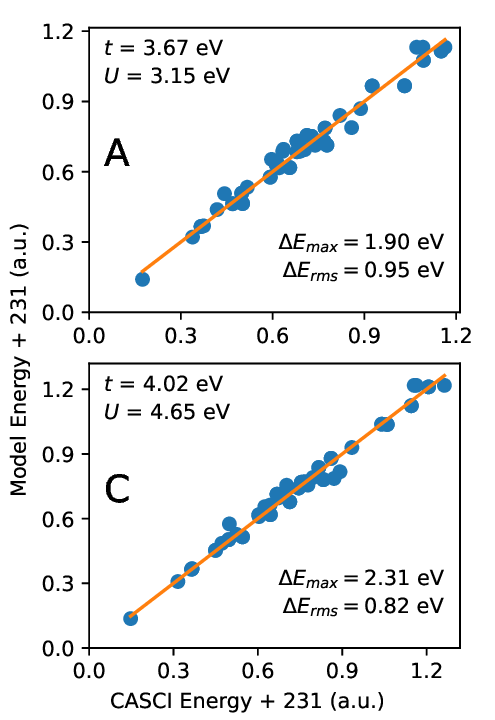}
  \caption{DMD for localized orbitals obtained for benzene via different applications of the PM method. }
  \label{fig:PMfit}
\end{figure}

Figure \ref{fig:PMfit} shows the final results of the DMD parameter fits for the benzene low-energy space using options A and C. Both fits appear visually reasonable, and the root mean square (RMS) errors are relatively similar. Therefore, the quality of the fit alone should not be used as an indicator of the quality of a model Hamiltonian without first considering the choice of orbitals.

\begin{table}[h!]
\caption{CASCI ground state energy and Hubbard model parameters for naphthalene, obtained with the DMD and CASCI electronic states}
\begin{center}
\label{tab:Hub_C10H8}
\begin{tabular}{ lccc}
\hline
Option& Ground state energy, a.u.&$U$, eV & $t$, eV \\
\hline
A&-383.554522& 3.39& 5.77 \\
B&-383.554522& 3.23 & 5.80\\
C&-383.599391& 3.92 & 9.10 \\
D&-383.599392& 3.92 & 9.10 \\
\hline
\end{tabular}
\end{center}
\end{table}

Unlike in the case of benzene, options A and B do not produce identical results for naphthalene. The FB and PM methods are based on fundamentally different localization criteria (spatial distance-based functionals versus charge-dependent metrics), and they produce slightly different orbitals even when operating within the same small subspace of orbitals. 
Comparing options A and C for naphthalene, the CASCI energy difference is approximately 1.2 eV (Table \ref{tab:Hub_C10H8}), and the differences in the extracted on-site Coulomb repulsion and hopping integrals are approximately 0.5 eV and 3.3 eV, respectively (Table \ref{tab:Hub_C10H8}). As in the case of benzene, the comparison between options A and C demonstrate that differences in the orbital subspace used to construct localized orbitals can lead to significant variations in the resulting Hubbard parameters. Such differences are expected to become even more significant for larger molecular systems or systems requiring larger active spaces, where the number of possible orbital choices increases.

\subsection*{\ce{FeSe}}
\label{sec:tmcs}
We consider the FeSe molecule as a simple example of a transition metal complex, for which a more extended model Hamiltonian is needed.
Its small size allows us to examine many orbital choices and clearly demonstrate the effects of different localization procedures.
For this molecule, we obtained the SCF wavefunction from a Kohn-Sham DFT calculations using the def2-TZVP basis set and the PBE0 functional \cite{adamo1999toward}. 
We computed 60 electronic states in total, corresponding to 20 states for each value of S=0, 1 and 2.

Considering a model orbital space consisting of the Fe $3d$ and $4s$ valence orbitals and the Se $4p$ orbitals, the model Hamiltonian in Eq.~(\ref{eq:MH3}) contains the following parameters:

\begin{itemize}
   \item  zeroth-order (core) energy, $E_0$;
   \item  symmetry-equivalent orbital energies, $\epsilon_{d_{z^2}}$, $\epsilon_{d_{xz}}=\epsilon_{d_{yz}}=\epsilon_{d_{\pi}}$,  $\epsilon_{d_{xy}}=\epsilon_{d_{x^2-y^2}}=\epsilon_{d_{\sigma}}$, $\epsilon_{4s}$, $\epsilon_{p_z}$, $\epsilon_{p_x}=\epsilon_{p_y}=\epsilon_{p_\pi}$
   \item	the symmetry-allowed hopping terms, $t_{\sigma,pd}$  (interaction between $p_z$ and $d_{z^2}$-like orbitals), $t_{\sigma,ps}$ (interaction between $p_z$ and $4s$-like orbitals), and $t_\pi$ ($p_x-d_{xz}$ and $p_y-d_{yz}$ interactions);
   \item	the self-repulsion terms, $U_p$, and $U_d$;
   \item	the Coulomb interactions, $V_{pd}$, and $V_{ps}$;
   \item	Hund coupling between $d$ orbitals $d_i$ and $d_j$, $K_{ij}$.
\end{itemize}

 Here, we refer to the subspace spanned by orbitals with the targeted character ($3d$ and $4s$ atomic orbitals of the Fe atom and the $4p$ atomic orbitals of the Se atom)as the  ``model orbital space''. When deriving the model Hamiltonian parameters for FeSe, we examined several chemically motivated choices of the ``model orbital space'':
 
\begin{itemize}
\item A: PM applied to the model orbital space from a SCF calculation
\item B: FB applied to the model orbital space  from a SCF calculation
\item C: NPL applied to the model orbital space from a SCF calculation
\item D: PM applied to the model orbital space constructed using the AVAS methods
\item E: IAOs followed by construction of the model orbital space
\item F: PM performed separately on occupied molecular orbitals and and first five virtual molecular orbitals.
\item  G: NPL performed separately on occupied molecular orbitals and and first five virtual molecular orbitals.
 \end{itemize}
Last two options were included because localization of only the occupied space is insufficient for constructing a complete valence model orbital space, while localization of the entire virtual space is generally problematic.

\begin{figure}[h!]
  \centering
  \includegraphics[width=0.9\columnwidth]{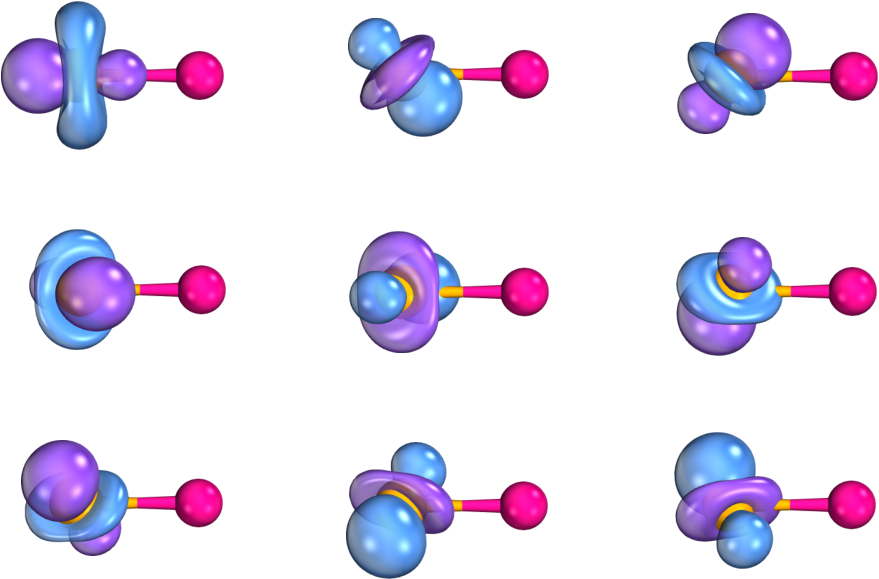}
  \caption{Localized orbitals obtained using the FB method on the entire space of occupied canonical orbitals from the reference SCF calculation for FeSe. This localization yields nine hybridized molecular orbitals, each containing a mixture of Fe  $3s$, $3p$ and $3d$ atomic orbitals character. The orbitals are visualized with the isosurface threshold corresponding to a 95$\%$ density cutoff.}
  \label{fig:Boys}
\end{figure}

As it was mentioned above, the ER and FB localization techniques, which use localization functionals depending only on spatial distances, typically mix atomic orbitals within the same shell. For example, the FB method mixes $3s$, $3p$ and $3d$ atomic orbitals in localized molecular orbitals when it is applied to the entire space of occupied orbitals of FeSe.  
Figure \ref{fig:Boys} shows  nine hybridized localized orbitals that can be obtained by localizing all occupied orbitals for FeSe using the FB method. These orbitals are intentionally visualized using a 95 $\%$ orbital density cutoff for clarity. 
Such hybridized orbitals are not optimal for describing electron $d \rightarrow d$ transitions.
However, one can obtain localized molecular orbitals with predominantly $3d$ character using FB and ER methods by applying their localization functional to a (pre-selected) set of SCF canonical orbitals that exclude molecular orbitals primarily representing the Fe $3s$ and $3p$ character.

As discussed above, localization of the complete virtual orbital space is generally not appropriate.
Moreover, such localization is usually unnecessary because model Hamiltonians typically require only a small number of low-energy virtual orbitals (nonbonding or antibonding) derived from the valence atomic orbitals.
Therefore, practical strategies usually involve either (i) localizing a small subset of low-energy virtual orbitals selected according to their energy or chemical character, or (ii) including a small number of virtual orbitals with the desired character together with the occupied orbitals during localization.
In the former approach, the quality of the resulting localized orbitals depends on the number and choice of virtual orbitals included in the localization subspace. 
Since orbital localization is performed through an iterative sequence of 2 x 2 unitary orbital rotations, the final localized orbitals can depend strongly on the orbital subspace provided to the localization algorithm. In particular, including or excluding orbitals with similar atomic orbital composition may significantly change the final localization.

\begin{figure}[h!]
  \centering
  \includegraphics[width=0.9\columnwidth]{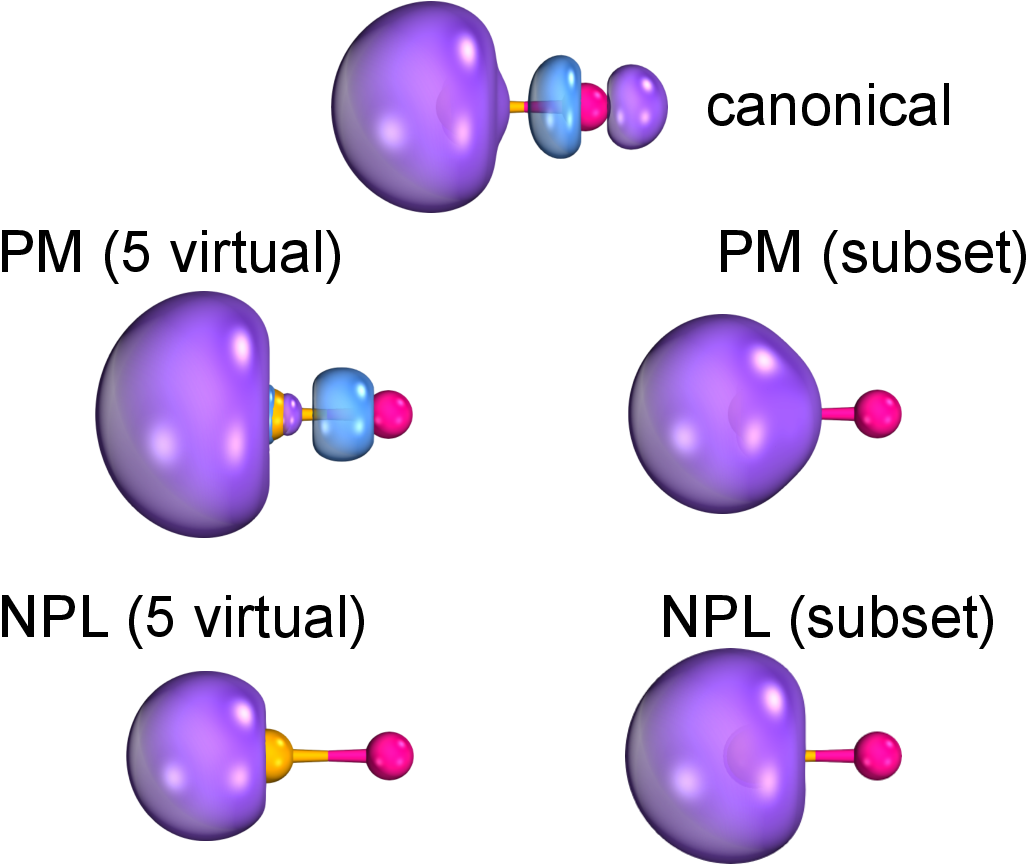}
  \caption{Localized $4s$ orbitals obtained for FeSe using different options. The orbitals are visualized with the isosurface threshold corresponding to an 80$\%$ density cutoff.}
  \label{fig:Orb4s}
\end{figure}

These effects are illustrated below using the example of a single 4s orbital of the iron atom in FeSe. Figure \ref{fig:Orb4s} compares the canonical orbital with localized molecular orbitals having predominantly Fe 4s atomic orbital character obtained using the PM and NPL functionals.
The localized orbitals were obtained using two different localization spaces. In the first approach, localization was performed on  the model orbital space from a SCF calculation  (Options A and C), i.e. a subspace consisting of eight occupied and one virtual canonical orbital carrying the targeted atomic orbital character (Fe 3d and 4s, together with Se 4p). In the second, only the five lowest-energy virtual canonical orbitals were localized (Options F and G). 
This comparison demonstrates how the same localization functional can produce noticeably different orbitals depending solely on the choice of localization subspace.  Such differences are expected to become even more pronounced in larger and more complex systems, than FeSe and can  lead to appreciable changes in the resulting model Hamiltonian parameters.

 In Figure \ref{fig:Orb4s} the orbitals are visualized with the 80$\%$ density cutoff, sufficient for visually observing the differences. The differences become even more apparent when a higher density cutoff is used for visualization.

Now let us examine the CASCI ground state energies together with the resulting model Hamiltonian parameters obtained using nine localized orbitals constructed with different localization options.
The ground state energies for options A-C are practically identical (Table \ref{tab:DMD_FeSe}), as these orbitals are obtained by applying different localization functionals to the same pre-selected model orbital space consisting of eight occupied and one virtual canonical orbital.
However, the excited-state energies differ slightly because each localization functional produces somewhat different localized orbitals.
Although these energy differences are small, they accumulate during the DMD fitting procedure. Small variations in the density matrix elements across many electronic states therefore lead to significant differences in the model Hamiltonian parameters (Table \ref{tab:DMD_FeSe}), reaching approximately 0.88 eV for the Coulomb interaction $V_{p-d}$ and 1.68 eV for the  $t_{\sigma,pd}$  hopping terms for the symmetry allowed interactions between $p$ and $d$ orbitals.

When the localized orbitals differ more significantly (for example, options A and E), as reflected by the larger differences in the CASCI ground-state energies, even larger variations in the model Hamiltonian parameters are obtained.
One can notice that model Hamiltonian parameters  differ not only in their magnitudes but, in some cases, even in their signs.
The relative importance of different physical interactions also changes; for example, the balance between hopping terms and Coulomb interactions can vary considerably.
Since FeSe is a relatively small molecule, even larger differences may be expected for larger systems with more extensive orbital spaces.
Such variations can ultimately lead to qualitatively different physical interpretations based on the resulting model Hamiltonians.

In this work, these RDMs are obtained from CASCI calculations and therefore describe only the static correlation captured within the active space. Ideally, the fitting should instead employ RDMs that also include dynamic correlation, which would provide a more accurate representation of the electronic structure and improve the quality of the resulting effective Hamiltonians. Although multireference perturbation theories such as NEVPT2 recover dynamic correlation accurately at the energy level, implementations capable of computing the corresponding one- and two-particle RDMs are currently not available. The development of dynamically correlated RDMs for DMD therefore remains an important direction for future work.
In addition, in many practical applications, NEVPT2 calculations for dynamic correlation become computationally too expensive because their cost scales approximately as the sixth power of the total number of orbitals. As a result, they are often impractical for large systems and the large basis sets required to accurately describe the electronic structure of strongly correlated systems.
This is the reason why in practice the choice of the localization technique for accurate treatment of static correlation is an important issue.

\section{Conclusions}

In this work, we have drawn attention to the problem of orbital choice when constructing model Hamiltonians from \textit{ab initio} calculations. We have shown that the resulting model Hamiltonian parameters depend not only on the localization functional itself, but also, and often more importantly, on the choice of the orbital space to which the localization procedure is applied. Even when different orbital choices produce nearly identical total electronic energies and similarly accurate fits to the low-energy spectrum, they can lead to significantly different model Hamiltonian parameters. These differences become even more pronounced when different model orbital spaces are employed and are expected to increase further for larger molecular systems and larger orbital spaces. As a consequence, different orbital choices may lead to qualitatively different physical interpretations based on the resulting model Hamiltonians. Therefore, the choice of model orbital space and the details of the localization procedure should be regarded as an integral part of model Hamiltonian construction and should be reported explicitly to ensure meaningful comparisons and reproducibility.

\onecolumngrid\
\begin{table}[h!]
\caption{CASCI ground state (S=2) energy and model parameters for FeSe, obtained using different orbital sets for the FeSe molecule}
\begin{center}
\label{tab:DMD_FeSe}
\begin{tabular}{ lccccccccccc}
\hline
Localization option& GS energy, a.u.&$ E_0 $& K & $t_{p-d,\sigma}$ & $t_{p-s}$ &$ t_{p-d,\pi}$ &$ U_d$ & $V_{p-d} $&$ e_{d,\delta}$ &$ R^2$ & $\Delta E_{rms}$ \\
\hline
A&-3662.247429& -99653.97& -2.27& -1.18& -0.36& 0.07& 0.33& 0.08& -0.37& 0.9589& 0.23\\
B&-3662.247427& -99649.05& -1.80& 0.50& 0.17& 0.13& 0.59& -0.14& -0.29& 0.9575& 0.23\\
C&-3662.247429& -99644.66& -1.68& -0.15& 0.39& 0.17& 0.62& -0.31& -0.22& 0.9553& 0.24\\
D&-3662.203991&  -99651.13& -1.93& -1.99& -1.19& 0.05& 0.47& 0.03& -0.38& 0.9605& 0.19\\
E&-3662.344214&  -99633.90&-1.99& -0.28& 1.30& 0.63& 0.52& -0.80& 0.00& 0.9916& 0.11\\
\hline
\end{tabular}
\end{center}
\end{table}

\twocolumngrid


\section*{Acknowledgements}
We acknowledge the Pennsylvania State University for providing the startup fund that supported this research. 

\cleardoublepage
\bibliography{refs}

\end{document}